\documentclass[12pt]{article}
\usepackage{natbib,graphicx,setspace,amsmath,amssymb,subfigure,url}
\newcommand{\bay}{\begin{array}}
\newcommand{\eay}{\end{array}}

\newcommand{\bqa}{\begin{eqnarray*}}
\newcommand{\eqa}{\end{eqnarray*}}
\newcommand{\bqan}{\begin{eqnarray}}
\newcommand{\eqan}{\end{eqnarray}}
\newcommand{\bqt}{\begin{quote}}
\newcommand{\eqt}{\end{quote}}
\newcommand{\bt}{\begin{tabbing}}
\newcommand{\et}{\end{tabbing}}
\newcommand{\bit}{\begin{itemize}}
\newcommand{\eit}{\end{itemize}}
\newcommand{\ben}{\begin{enumerate}}
\newcommand{\een}{\end{enumerate}}
\newcommand{\beq}{\begin{equation}}
\newcommand{\eeq}{\end{equation}}
\newcommand{\bdefi}{\begin{definition}}
\newcommand{\edefi}{\end{definition}}
\newcommand{\bpro}{\begin{proposition}}
\newcommand{\epro}{\end{proposition}}
\newcommand{\bco}{\begin{corollary}}
\newcommand{\eco}{\end{corollary}}
\newcommand{\bdes}{\begin{description}}
\newcommand{\edes}{\end{description}}

\def\log{\hbox{log}}

\def\boxit#1{\vbox{\hrule\hbox{\vrule\kern6pt
          \vbox{\kern6pt#1\kern6pt}\kern6pt\vrule}\hrule}}

\def\bse{\begin{eqnarray*}}
\def\ese{\end{eqnarray*}}
\def\be{\begin{eqnarray}}
\def\ee{\end{eqnarray}}
\def\bq{\begin{equation}}
\def\eq{\end{equation}}

\def\T{{\cal T}}

\newtheorem{proposition}{Proposition}

\newcommand{\blem}{\begin{lemma}}
\newcommand{\elem}{\end{lemma}}
\newcommand{\bthe}{\begin{theorem}}
\newcommand{\ethe}{\end{theorem}}

\newtheorem{definition}{Definition}[section]
\newtheorem{lemma}[definition]{Lemma}
\newtheorem{theorem}[definition]{Theorem}

\def\delete#1{\iffalse #1 \fi}

\def\bse{\begin{eqnarray*}}
\def\ese{\end{eqnarray*}}
\def\bee{\begin{enumerate}}
\def\eee{\end{enumerate}}
\def\bqe{\begin{eqnarray}}
\def\eqe{\end{eqnarray}}
\def\bed{\begin{description}}
\def\eed{\end{description}}
\def\bei{\begin{itemize}}
\def\eei{\end{itemize}}

\def\pmb#1{\setbox0=\hbox{#1}%
    \kern-.025em\copy0\kern-\wd0
    \kern.05em\copy0\kern-\wd0
    \kern-.025em\raise.0433em\box0 }
\def\pmbh#1#2{\setbox0=\hbox{#1}%
    \setbox1=\hbox{#2}%
    \kern-.025em\copy0\kern-\wd0
    \kern.05em\copy1\kern-\wd0
    \kern-.025em\raise.0433em\box0 }

\def\frac#1#2{{#1\over#2}}

\def\boxit#1{\vbox{\hrule\hbox{\vrule\kern6pt
   \vbox{\kern6pt#1\kern6pt}\kern6pt\vrule}\hrule}}

\def\listing#1{\vskip 4mm\begin{verbatim}\input#1 \vskip 4mm}
\def\thick#1{\hbox{\rlap{$#1$}\kern0.25pt\rlap{$#1$}\kern0.25pt$#1$}}

\def\T{{\mbox{\rm\tiny T}}}

\def\pmbh{{\pmb h}}

\def\calA{{\cal A}}

\def\calC{{\cal C}}

\def\calF{{\cal F}}
\def\calG{{\cal G}}

\def\calW{{\cal W}}

\renewcommand\today{\ifcase\month\or
   Jan\or Feb\or Mar\or Apr\or May\or
   Jun\or Jul\or Aug\or Sep\or Oct\or Nov\or
   Dec\fi
   \space\number\day, \number\year}

\newtheorem{thm}{Theorem}

\newtheorem{rmk}{Remark}
\newtheorem{prp}{Proposition}

\begin{document}

\title{Minimax Prediction for Functional Linear Regression with Functional Responses in Reproducing Kernel Hilbert Spaces}         

\author{
Heng Lian\\
\begin{tabular}{c}
 {\small\it Division of Mathematical Sciences, School of Physical and Mathematical Sciences,}\\
 {\small\it  Nanyang Technological University, Singapore, 637371}
\end{tabular}} 
\date{}          
\maketitle
\begin{abstract}
In this article, we consider convergence rates in functional linear regression with functional responses, where the linear coefficient lies in a reproducing kernel Hilbert space (RKHS). Without assuming that the reproducing kernel and the covariate covariance kernel are aligned, or assuming polynomial rate of decay of the eigenvalues of the covariance kernel, convergence rates in prediction risk are established. The corresponding lower bound in rates is derived by reducing to the scalar response case. Simulation studies and two benchmark datasets are used to illustrate that the proposed approach can significantly outperform the functional PCA approach in prediction.

\noindent\textbf{keywords:}  Functional data; Functional response; Minimax convergence rate; Regularization.
\end{abstract}

\section{Introduction}
The literature contains an impressive range of functional analysis tools for various problems
including exploratory functional principal component analysis, canonical correlation
analysis, classification and regression. Two major approaches exist. The more traditional
approach, masterfully documented in the monograph \citep{ramsay05}, typically
starts by representing functional data by an expansion with respect to a certain basis, and
subsequent inferences are carried out on the coefficients. The most commonly utilized basis
include B-spline basis for nonperiodic data and Fourier basis for periodic data. Another line
of work by the French school \citep{ferraty02}, taking a nonparametric point of view, extends the traditional nonparametric techniques, most notably the kernel estimate, to the functional case. Some recent advances in the area of functional regression include \cite{cardot03,cai06,preda07,lian07,saidi08,yao05,crambes09,presmoothing11,lian11a}.

In this paper we study the functional linear regression problem of the form
\begin{equation}\label{eqn:model}
Y(t)=\mu(t)+\int_0^1\beta(t,s)X(s)\,ds+\epsilon(t),
\end{equation}
where $Y,X,\epsilon\in L_2[0,1]$ and $E[\epsilon|X]=0$, the same problem that appeared in \cite{ramsay05,yao05,antoch08,aguilera08,mas12}. In terms of methodology, the plan of attack we will give for (\ref{eqn:model}) is most closely related to that of \cite{mas12}. In this introduction, we will explain the methodology used in that paper and then the different assumption we will make on $\beta(t,s)$.

Without loss of much generality, throughout the paper we assume $E(X)=0$ and the intercept $\mu(t)=0$, since the intercept can be easily estimated. The covariance operator of $X$ is the linear operator $\Gamma=E(X\otimes X)$ where for $x,y\in L_2[0,1]$, $x\otimes y: L_2[0,1]\rightarrow L_2[0,1]$ is defined by $(x\otimes y)(g)=\langle y,g\rangle x$ for any $g\in L_2[0,1]$. $\Gamma$ can also be represented by the bivariate function $\Gamma(s,t)=E[X(s)X(t)]$. Using the same letter $\Gamma$ to denote both the operator and the bivariate function will not cause confusion in our context. We assume throughout the paper that $E\|X\|^4<\infty$ which implies $\Gamma$ is a compact operator. Then by the Karhunen-Lo\`eve Theorem there exists a spectral expansion for $\Gamma$,
\[\Gamma=\sum_{j=1}^\infty\lambda_j\varphi_j\otimes\varphi_j,\]
where $\lambda_j\ge 0$ are the eigenvalues with $\lambda_j\rightarrow 0$ and $\{\varphi_j\}$ are the orthonormalized eigenfunctions. Correspondingly, we have the representation $X=\sum_{j\ge 1}\gamma_j\varphi_j$ with $\gamma_j=\int X\varphi_j$. The random coefficients $\gamma_j$ satisfies $E\gamma_j\gamma_k=\lambda_jI\{j=k\}$ where $I\{.\}$ is the indicator function.

By expanding $\beta$ using the set of eigenfunctions, we write $\beta(t,s)=\sum_{j\ge 1}b_j(t)\varphi_j(s)$ and (\ref{eqn:model}) can be equivalently written as
\[Y(t)=\sum_{j\ge 1}b_j(t)\gamma_j+\epsilon(t).\]
Multiplying both sides above by $\gamma_j$ and taking expectations, we easily obtain $b_j(t)=E[Y(t)\gamma_j]/\lambda_j$. Given i.i.d. data $(X_i,Y_i), i=1,\ldots,n$, \{$\lambda_j$, $\varphi_j$\} can be easily estimated by $\hat{\lambda}_j$ and $\hat{\varphi}_j$ obtained from the spectral decomposition of the empirical covariance operator and $E[Y(t)\gamma_j]$ can be approximated by the corresponding sample average. Thus the estimator proposed in \cite{mas12} is 
\[\hat{\beta}(t,s)=\frac{1}{n}\sum_{i=1}^n\sum_{j=1}^k\frac{\int X_i\hat{\varphi}_j}{\hat{\lambda}_j}Y_i(t)\hat{\varphi}_j(s).\]
Note that the infinite sum over $j$ has been truncated as some point $k$ for regularization. One intriguing point is that there is no regularization on $Y_i(t)$ necessary, in contrast with \cite{yao05} where $Y$ is observed sparsely with additional noise. This can also be seen from that $b_j(t)$ is not a priori constrained in any way. The reason is that only regularization of the covariance operator, which does not depend on $Y$, is necessary to avoid overfitting. 

Minimax convergence rates of $E\|\int \hat{\beta}(t,s)X(s)ds-\int\beta(t,s)X(s)ds\|^2$ were shown in \cite{mas12}. A key assumption is the appropriate decaying assumption on $\|b_j\|$ as $j$ increases. Given that $\|b_j\|$'s are the coefficients of $\beta(t,s)$ in terms of the basis $\varphi_j$, which is a characteristic of the predictor, there is no a priori reason why this basis should provide a good representation of $\beta$ in the sense that $\|b_j\|$ will decay fast. Indeed, a more reasonable assumption for $\beta$ is on its smoothness, which makes a reproducing kernel Hilbert space (RKHS) approach more reasonable conceptually. Such arguments have led to the developments in \cite{yuancai10,caiyuan12} for the scalar response models. While \cite{mas12} is based on \cite{cardot07} for scalar response models, ours is based on \cite{caiyuan12}. 

The rest of the article is organized as follows. In Section 2, we propose an estimator for $\beta$ with an RKHS approach where the reproducing kernel and the covariance kernel are not necessarily aligned. We establish the minimax rate of convergence in prediction risk by deriving both the upper bound and the lower bound. In Section 3, we present some simulation studies to show that the RKHS approach could significantly outperform the functional PCA approach when the kernels are mis-aligned. This advantage is further illustrated on two benchmark datasets which shows better prediction performance using our approach. We conclude in Section 4 with some discussions. The technical proofs are relegated to the Appendix.

Finally, we list some notations and properties regarding different norms to be used. For any operator $\calF$, we use $\calF^\T$ to denote its adjoint operator. If $\calF$ is self-adjoint and nonnegative definite, $\calF^{1/2}$ is its square-root satisfying $\calF^{1/2}\calF^{1/2}=\calF$. For $f\in L_2$, $\|f\|$ denotes its $L_2$ norm. For any operator $\calF$, $\|\calF\|_{op}$ is the operator norm $\|\calF\|_{op}:=\sup_{\|f\|\le 1}\|\calF f\|$. The trace norm of an operator $\calF$ is ${\rm Trace}(\calF)=\sum_k\langle(\calF^\T \calF)^{1/2}e_k,e_k\rangle$ for any orthonormal basis $\{e_k\}$ of $L_2$. $\calF$ is a trace class operator if its trace norm is finite. The Hilbert-Schmidt norm of an operator is $\|\calF\|_{HS}=(\sum_{j,k}\langle \calF e_j,e_k\rangle^2)^{1/2}=(\sum_{j}\| \calF e_j\|^2)^{1/2}$. An operator is a Hilbert-Schmidt operator if its Hilbert-Schmidt norm is finite. From the definition it is easy to see that ${\rm Trace}(\calF^\T\calF)={\rm Trace}(\calF\calF^\T)=\|\calF\|_{HS}^2$, Furthermore, if $\calF$ is a Hilbert-Schmidt operator and $\calG$ is a bounded operator, then $\calF\calG$ is also a Hilbert-Schmidt operator with $\|\calF\calG\|_{HS}\le \|\calF\|_{HS}\|\calG\|_{op}$.

\section{Methodology and Convergence Rates}

Following \cite{wahba90}, a RKHS ${H}$ is a Hilbert space of real-valued functions defined on, say, the interval $[0,1]$, in which the point evaluation operator $L_t: H\rightarrow R, L_t(f)=f(t)$ is continuous. By Riesz representation theorem, this definition implies the existence of a bivariate function $K(s,t)$ such that
\begin{eqnarray*}\label{rep}
&&K(s,\cdot)\in H, \mbox{ for all } s\in [0,1]\nonumber\\
&&\mbox{and (reproducing property)} \nonumber \\
&&\mbox{for every } f\in H \mbox{ and } t\in [0,1],\;  \langle K(t,\cdot), f\rangle_H=f(t).
\end{eqnarray*}
The definition of a RKHS can actually start from a positive definite bivariate function $K(s,t)$ and RKHS is constructed as the completion of the linear span of $\{K(s,\cdot),s\in [0,1]\}$   with inner product defined by $\langle K(s,\cdot),K(t,\cdot)\rangle_H=K(s,t)$. To make the dependence on $K$ explicit, the RKHS is denoted by $H_K$ with the RKHS norm $\|\cdot\|_{H_K}$. With abuse of notation, $K$ also denotes the linear operator $f\in L_2\rightarrow Kf=\int K(\cdot,s)f(s)ds$. For later use, we note that $H_K$ is identical to the range of $K^{1/2}$. 

We assume that for any $t\in[0,1]$, $\beta(t,\cdot)\in H_K$. This is a smoothness assumption for $\beta(t,s)$ in the $s$-variable. As noted in the introduction, smoothness assumption on the $t$-variable is not necessary. We estimate $\beta$ via
\begin{equation}\label{eqn:min}
\hat{\beta}=\arg\min_{\beta(t,.)\in H_K}\frac{1}{n}\sum_{i=1}^n\|Y_i-\int_0^1\beta(\cdot,s)X_i(s)\,ds\|^2+\lambda \int_0^1 \|\beta(t,\cdot)\|_{H_K}^2 dt.
\end{equation}
We implicitly assume that the expression $\int_0^1 \|\beta(t,\cdot)\|_{H_K}^2 dt$ is valid, that is $\|\beta(t,\cdot)\|_{H_K}$ as a function of $t$ is square integrable. This assumption on $\beta$ is also more succinctly denoted by $\beta\in L^2\times H_K$.

The following representer theorem is useful in computing the solution, whose proof is omitted since it is standard.

\begin{prp}
The solution of (\ref{eqn:min}) can be expressed as 
\begin{equation}\label{eqn:representer}
\hat{\beta}(t,s)=\sum_{i=1}^n c_i(t)\int_0^1 K(s,u)X_i(u)\,du.
\end{equation}
\end{prp}

Based on the previous proposition, by plugging the representation (\ref{eqn:representer}) into (\ref{eqn:min}), it can be easily shown that $(c_1(t),\ldots,c_n(t))^\T=(\Sigma+n\lambda)^{-1}Y(t)$ where $\Sigma$ is an $n\times n$ matrix whose entries are given by $\Sigma_{ij}=\int\int X_i(s)K(s,t)X_j(t)dsdt$.

\begin{rmk}\label{rmk:1} Throughout this section, we assume the reproducing kernel $K$ is positive definite and the RKHS norm for $H_K$ is used in the penalty. More generally, for practical use, we can assume $H_K=H_1\oplus H_2$, where $H_1$, typically finite dimensional, is a RKHS with reproducing kernel $K_1$ and $H_2$ is a RKHS with reproducing kernel $K_2$, $K=K_1+K_2$. We can then impose the penalty $\int \|P_2\beta(t,\cdot)\|^2_{H_K}dt=\int \|P_2\beta(t,\cdot)\|^2_{H_2}dt$, where $P_2$ is the projection onto $H_2$. Our theory and computation can be easily adapted to this more general case, but we use (\ref{eqn:min}) for ease for presentation throughout the paper. In real data analysis, $H_K=\calW_2^{per}$ is the second-order Sobolev space of periodic functions on $[0,1]$ and we use decomposition $H_K=H_1\oplus H_2$ where $H_1$ contains the constant functions.
\end{rmk}

Since $\beta(t,\cdot)\in H_K$, there exists $f(t,s)$ such that $\beta(t,\cdot)=K^{1/2}f(t,\cdot)$ and $\|\beta(t,\cdot)\|_{H_K}=\|f(t,\cdot)\|$. Thus (\ref{eqn:min}) can also be written as
\begin{equation}\label{eqn:min2}
\hat{f}=\arg\min_{f\in L_2[0,1]^2}\frac{1}{n}\sum_{i=1}^n\|Y_i-\int_0^1f(\cdot,s) (K^{1/2}X_i)(s)\,ds\|^2+\lambda \int_0^1\int_0^1 f^2(t,s) dsdt.
\end{equation}
Due to the appearance of $K^{1/2}X_i$ in the expression above, this suggests that the spectral decomposition of $T:=K^{1/2}\Gamma K^{1/2}$ plays an important role. Suppose the spectral decomposition of $T$ is 
\[T=\sum_{j\ge 1}s_j e_j\otimes e_j,\]
with $s_1>s_2>\cdots>0$. 

The following technical assumptions are imposed.
\begin{itemize}
\item[(A1)] There exists a positive, convex, decreasing function $\phi: (0,\infty)\rightarrow R^+$ such that $s_j=\phi(j)$ at least for large $j$. 
\item[(A2)] Recall the Karhunen-Lo\'eve expansion $K^{1/2}X=\sum_{j\ge 1}\xi_je_j$. There exists a constant $c$ such that $E[\xi^4_j]\le c(E[\xi^2_j])^2$ for all $j\ge 1$.
\item[(A3)] $\beta(t,\cdot)\in H_K$ for all $t\in [0,1]$, and $\|\beta(t,\cdot)\|_{H_K}\in L_2$ as a function of $t$. Furthermore, $BK^{-1/2}$ is a Hilbert-Schmidt operator, where the operator $B$ is defined by $Bf=\int \beta(\cdot,s)f(s)ds, f\in L_2$.
\end{itemize}

Assumption (A1) also appeared in \cite{cardot07}. \cite{caiyuan12} considered a much more restrictive polynomial decay assumption $s_j\asymp j^{-2r}$ for some $r>0$, which corresponds to $\phi(x)=x^{-2r}$. Taking $\phi(x)=c_1e^{-c_2x}$  for some constants $c_1,c_2>0$, exponential decay of eigenvalues is also a special case of our result, among many others. 

Assumption (A2) is similar to that assumed in \cite{hall07,cardot07}. \cite{caiyuan12} assumed that $E(\int X(t)f(t)dt)^4\le c(E(\int X(t)f(t)dt)^2)^2$ for all $f\in L_2$. This assumption implies (A2) which can be seen by choosing $f=K^{1/2}e_j$.

(A3) is a natural extension of the case with scalar reponse, where $\beta(t)\in H_K$ automatically implies $K^{-1/2}\beta\in L_2$. Superficially, $BK^{-1/2}$ in (A3) is only defined on the range of $K^{1/2}$, which coincides with $H_K$ and is a dense subset of $L_2$. Also, since $K^{-1/2}$ is an unbounded operator, it is not clear that $BK^{-1/2}$ can be bounded. Nevertheless, it can be shown that under the condition that $\beta(t,\cdot)\in H_K$ and $\|\beta(t,\cdot)\|_{H_K}\in L_2$, $BK^{-1/2}$ is bounded on $L_2$. More specifically, we have the following proposition whose proof is in the Appendix.
\begin{prp}\label{prp:close}
If $\beta(t,\cdot)\in H_K$ for all $t\in [0,1]$ and $\|\beta(t,\cdot)\|_{H_K}\in L_2$ where $\|\beta(t,\cdot)\|_{H_K}$ is regarded as a function of $t$, then $BK^{-1/2}$ is a bounded operator on $L_2$.
\end{prp}

The risk we consider is the prediction risk $E^*\|\hat{B}(X^*)-{B}(X^*)\|^2$ where $X^*$ is a copy of $X$ independent of the training data and $E^*$ is the expectation taken over $X^*$. We first present the upper bound.
\begin{thm}\label{thm:upper}
Under assumptions (A1)-(A3), and that $\lambda\rightarrow 0, \lambda n\rightarrow\infty$, we have 
$$E^*\|\hat{B}(X^*)-{B}(X^*)\|^2=O_p\left(\lambda+\frac{1}{n}\sum_j\frac{s_j^2}{(s_j+\lambda)^2}\right).$$
\end{thm}
\begin{rmk} By examining the proof carefully, one can actually see that the convergence is uniform in $\beta$ that satisfies (A3) with $\|BK^{-1/2}\|_{HS}\le 1$ (there is nothing special about the upper bound 1, which can be replace by any $L>0$). We can thus actually show
\[\lim_{a\rightarrow \infty}\lim_{n\rightarrow\infty}\sup_{\beta\in L_2\times H_K, \|BK^{-1/2}\|_{HS}\le 1}P(E^*\|\hat{B}(X^*)-{B}(X^*)\|^2>a\lambda_0)=0\]
This expression is put here for easy comparison with the lower bound obtained in Theorem \ref{thm:lower} below.
\end{rmk}

We now discuss how to choose appropriate $\lambda$ to balance the two terms in the rate above. Let $J=\lfloor \phi^{-1}(\lambda)\rfloor$ be the integer part of $\phi^{-1}(\lambda)$. By splitting the sum over $j$ into $j\le J$ and $j>J$, we have $$\frac{1}{n}\sum_j\frac{s_j^2}{(s_j+\lambda)^2}\le \frac{J}{n}+\frac{s_{J+1}\sum_{j\ge J+1}s_j}{n\lambda^2}.$$ Let $\lambda_0$ be the solution to the equation
\begin{equation}\label{eqn:lambda0}
\phi^{-1}(\lambda)=n\lambda.
\end{equation}
Then we have $J_0:=\lfloor\phi^{-1}(\lambda_0)\rfloor\le \phi^{-1}(\lambda_0)$ and $$\frac{s_{J_0+1}\sum_{j\ge J_0+1}s_j}{n\lambda_0^2}\le \frac{(J_0+2)s_{J_0+1}^2}{n\lambda_0^2}\le \frac{J_0+2}{n},$$ where we used that $\sum_{j\ge J_0+1}s_j\le (J_0+2)s_{J_0+1}$ obtained from Lemma 1 of \cite{cardot07}, and that $s_{J_0+2}=\phi(J_0+2)\le\phi(\phi^{-1}(\lambda_0))=\lambda_0$ by the definition of $J_0$. Thus we have
$$E^*\|\hat{B}(X^*)-\hat{B}(X^*)\|^2=O_p(\lambda_0)$$
with $\lambda_0$ defined by (\ref{eqn:lambda0}), which characterizes the optimal convergence rate. In the special case $\phi(x)=x^{-2r}$, $\lambda_0=n^{-2r/(2r+1)}$, which is the same as the rate obtained in \cite{caiyuan12} for scalar response models. On the other hand, if $\phi(x)=e^{-x}$, we can easily show that $\log\log n/n<\lambda_0<\log n/n$, an almost parametric rate.

We now establish the lower bound. This is obtained by first reducing the problem to the scalar response model and then using a slightly different construction from that used in \cite{caiyuan12} to deal with more general $\phi$. The details of the proof are contained in the Appendix.
\begin{thm}\label{thm:lower}
Under assumptions (A1) and (A2) on the predictor distribution, we have, for any $a>0$
\[\lim_{a\rightarrow 0}\lim_{n\rightarrow\infty}\inf_{\hat\beta}\sup_{\beta\in L_2\times H_K, \|BK^{-1/2}\|_{HS}\le 1}P(E^*\|\hat{B}(X^*)-{B}(X^*)\|^2>a\lambda_0)=1\]
where the infimum is taken over all possible estimators based on the training data $(X_i,Y_i),i=1,\ldots, n$.
\end{thm}

\section{Numerical Results}
\subsection{Simulations}
The simulation setup is similar to that used in \cite{caiyuan12}. We consider the RKHS with kernel
\[K(s,t)=\sum_{j\ge 1}\frac{2}{(j\pi)^4}\cos(j\pi s )\cos(j\pi t ),\]
and thus $H_K$ consists of functions of the form
\[f(t)=\sum_{j\ge 1}f_j\cos(j\pi t)\]
such that $\sum_j j^4f_j^2<\infty$. In this case, we actually have $\|f\|_{H_K}^2=\int (f'')^2$. Data are generated from (\ref{eqn:model}) without the intercept term, with 
$$\beta(t,s)=\sum_j4\sqrt{2}(-1)^j\frac{\sin(j\pi t)}{j^{2}}\cos(j\pi s).$$
For the covariance kernel, we use
$$\Gamma(s,t)=\sum_{j\ge 1}2\theta_j\cos(j\pi s)\cos(j\pi t),$$
where $\theta_j=(|j-j_0|+1)^{-2}$. When $j_0=1$, the two kernels are perfectly aligned, in the sense that they have the same sequence of eigenfunctions when ordered according to the eigenvalues. As $j_0$ increases, the level of mis-alignment also increases and we expect that the performance of functional PCA approach deteriorate with $j_0$. After finding the integral $Z_i(t):=\int \beta(t,s)X_i(s)ds$ (approximated easily by a Riemannian sum), we discretize $Z_i$ over $[0,1]$ on an equally-spaced grid $(t_1,\ldots,t_{100})$ with 100 points and then add independent $\epsilon_{ik}\sim N(0,\sigma^2)$ noises to finally obtain $Y_i(t_k)=Z_i(t_k)+\epsilon_{ik}$. The discretized data for model fitting contains $(X_i(t_k),Y_i(t_k)),k=1,\ldots,100, i=1,\ldots,n$. We set $n=50,100$ and $\sigma=0.1,0.3$, resulting in a total of four scenarios for each $j_0$. For values of $j_0$, we use $j_0\in \{1,3,5,\ldots,15\}$. For the functional PCA approach, the tuning parameter is the truncation point which we consider in the range from $1$ to $25$. For the RKHS approach, the tuning parameter is $\lambda$ and we consider $\lambda\in \exp\{-20,-19,\ldots,0\}$. The experiment for each scenario was repeated 100 times.

In this simulation, the tuning parameters are chosen to yield the smallest error to reflect the best achievable performance for both methods. To assess the performance, 100 test predictors $X_1^*,\ldots,X_{100}^*$ are generated from the same model as the training data, and root mean squared error (RMSE) is defined to be $(\sum_i\|\int\hat{\beta}X_i^*-\int\beta X_i^*\|^2/100)^{-1/2}$. Simulation results are summarized in Figure \ref{fig:sim}, which shows the RMSE for both methods. Each panel corresponds to a pair of values of $(n,\sigma)$, and the curves show the RMSE averaged over 100 replications for both methods as $j_0$ increases (red curve for the functional PCA approach and black curve for the RKHS approach). The vertical bar shows $\pm$ 2 standard errors computed from the 100 replications.

It is clearly seen that the performance of the RKHS approach is similar  to (actually better than) that of the functional PCA approach for $j_0=1$. As $j_0$ increases, the performance of the functional PCA approach becomes much worse, while the errors for the RKHS approach remain at the same level. The difference in performance between these two methods generally increases with $j_0$ (with some exceptions in our particular simulations).

\begin{figure}
\centerline{\includegraphics[width=5in,height=5in]{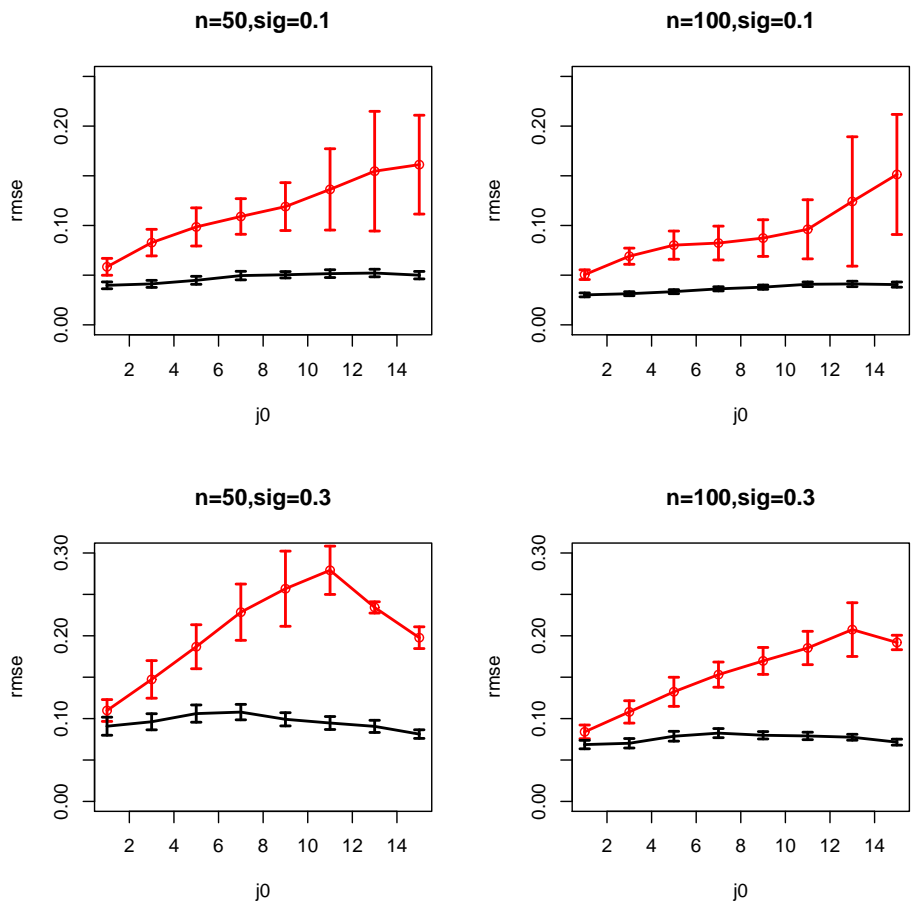}}
\caption{RMSE for both the functional PCA method (red curve) and the RKHS method (black curve) for the simulated data using the optimal tuning parameters. \label{fig:sim}}
\end{figure}

\subsection{Real data}
We now turn to the prediction performance of the proposed method on two real datasets. These datasets are used frequently in functional data analysis, and both are available from the fda package in R.

\textbf{Canadian weather data.} The daily weather data consists of daily temperature and precipitation measurements recorded in 35 Canadian weather stations. Each observation consists of functional data observed on an equally-spaced grid of 365 points. We treat the temperature as the independent variable and the goal is to predict the corresponding precipitation curve given the temperature measurements. As is previously done, we set the dependent variable to be the log-transformed precipitation measurements, and a small
positive number is added to the values with 0 precipitation recorded. Given the periodic nature of the data, we set $H_K=\calW_2^{per}$, the second-order Sobolev space of periodic functions on $[0,1]$. The reproducing kernel is given by $K(s,t)=K_1(s,t)+K_2(s,t)$ with $K_1(s,t)=1, K_2(s,t)=\sum_{j\ge 1}\frac{2}{(2\pi j)^4}\cos(2\pi j(s-t))$. The modification as mentioned in Remark \ref{rmk:1} is used. We use leave-one-out cross-validation to determine the best tuning parameters to use for both methods. The left panel in Figure \ref{fig:real} shows the prediction errors on the 35 stations using the best tuning parameters (trained on 34 stations). For 20 stations, the functional PCA approach has larger error than the RKHS approach. The average mean prediction error for the functional PCA approach is 0.43 while the error is 0.40 for the RKHS approach.

\textbf{Gait data.}  The Motion Analysis Laboratory at Children's Hospital, San Diego, collected these data, which consist of the angles formed by the hip and knee of 39 children over each child's gait cycle.  The cycle begins and ends at the point where the heel of the limb under observation strikes the ground. Both sets of functions are periodic and it is of interest to see how the two joints interact. In this application, we use hip angle as the predictor and knee angle as the response. The right panel in Figure \ref{fig:real} shows the prediction errors on the 39 children.  For 21 children, the functional PCA approach has larger error than the RKHS approach. The average mean prediction error for the functional PCA approach is 4.49 while the error is 4.38 for the RKHS approach.

\begin{figure}
\centering{
\includegraphics[width=2.5in,height=2.5in,trim=0 0 0 0,clip=TRUE]{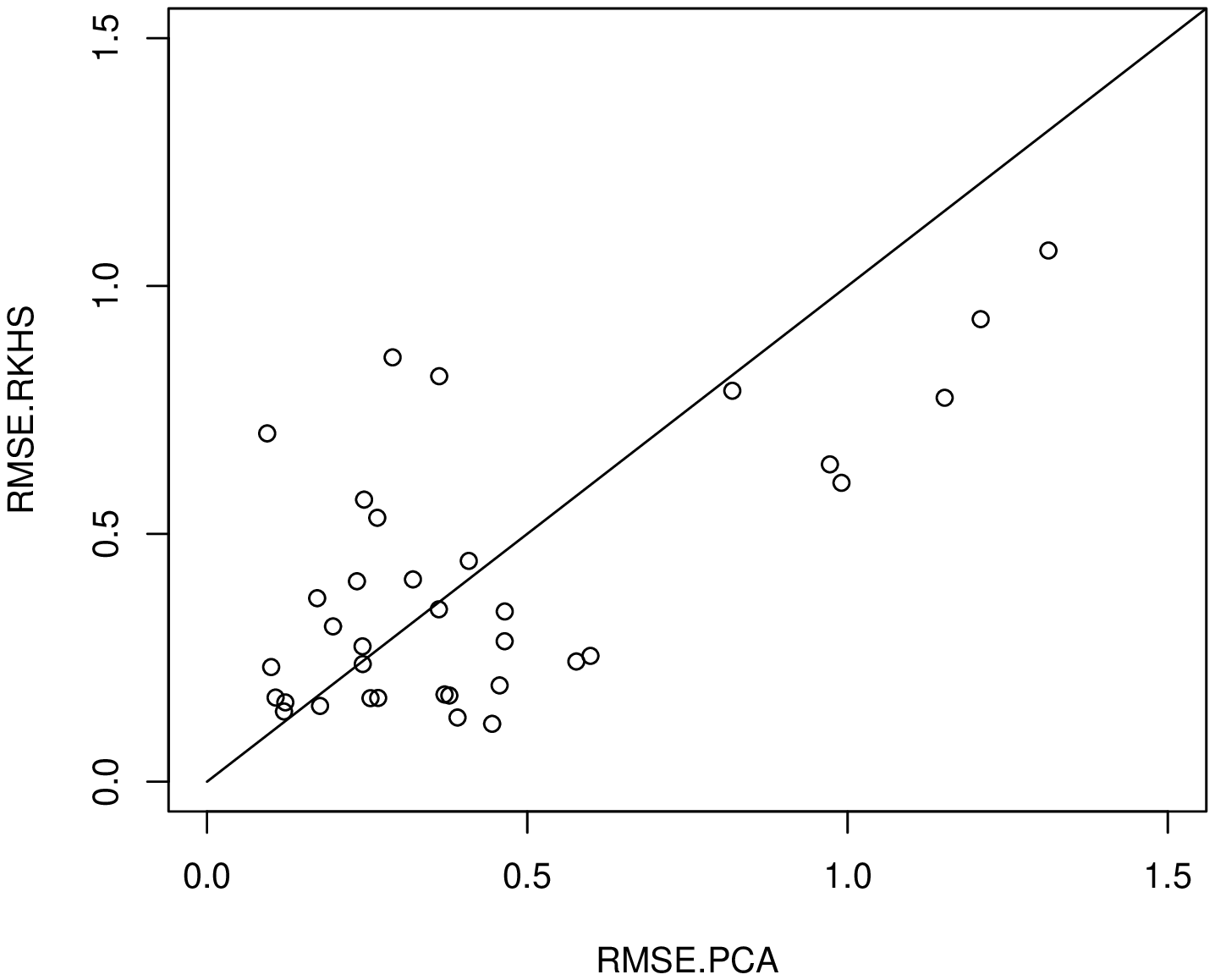}
\hfil
\includegraphics[width=2.5in,height=2.5in,trim=0 0 0 0,clip=TRUE]{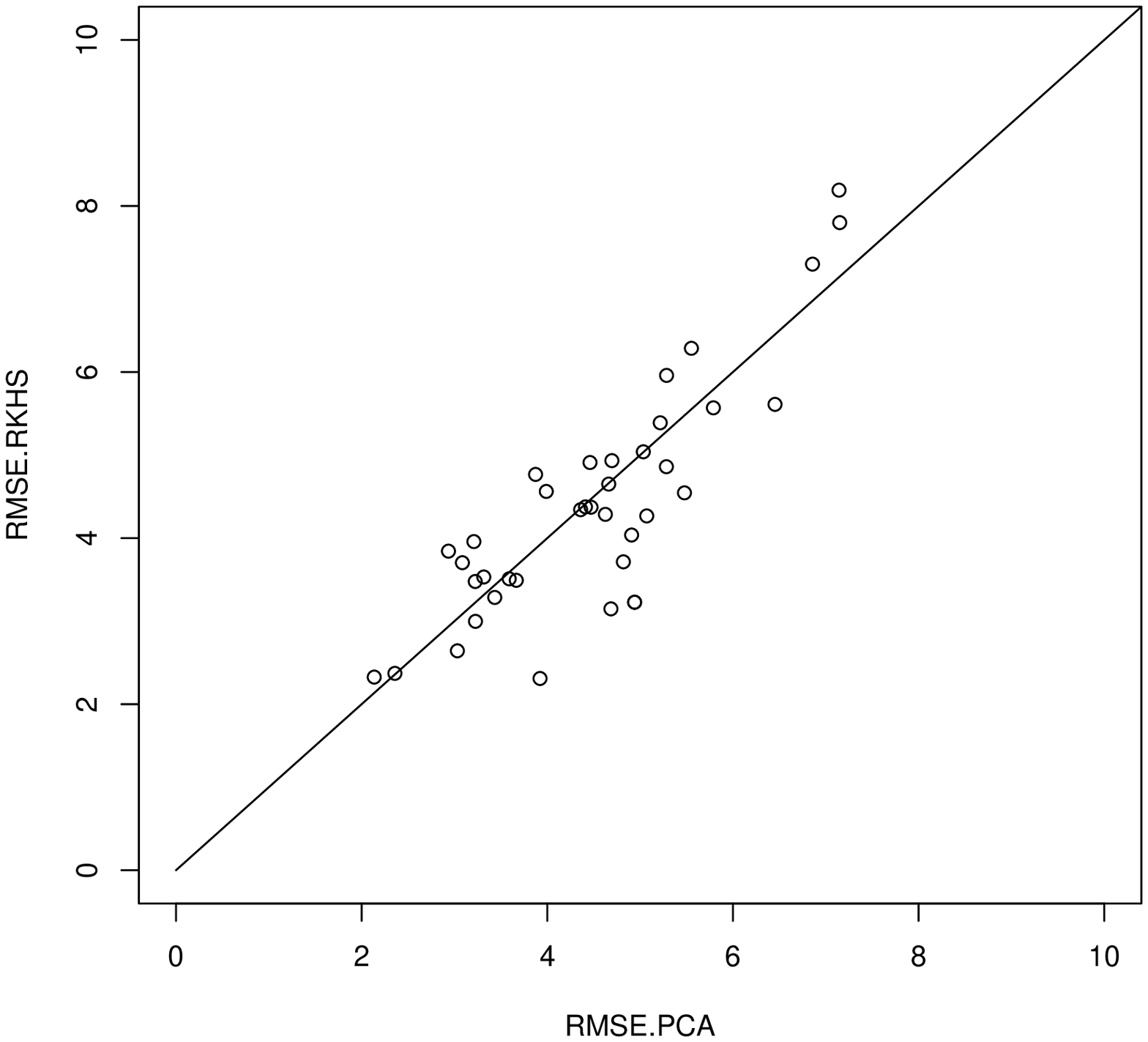}
}
\caption{Leave-one-out prediction error for the real data. The x-coordinate for each point shows the error of the functional PCA method, and the y-coordinate shows the error of the RKHS method. Left panel: Canadian weather data; Right panel: Gait data. The tuning parameters are chosen to minimize the leave-one-out cross-validation error in both methods. \label{fig:real}}
\end{figure}

\section{Conclusion}
In this paper, we established the minimax rate of convergence for prediction in functional response models in the general setting where the covariance kernel $\Gamma$ and the reproducing kernel $K$ are not aligned, and also under general assumption on the decay rate of the eigenvalues of operator $T=K^{1/2}\Gamma K^{1/2}$. Our simulations show that as the degree of alignment of the two kernels decreases, the RKHS estimator can significantly outperform the estimator based on functional PCA. The two real datasets further demonstrate that the RKHS estimator can have better prediction accuracy.

Choice of tuning parameter $\lambda$ can be done via cross-validation, as illustrated in our analysis of the real data. \cite{caiyuan12} proposed an adaptive method for tuning parameter selection which is an important theoretical development, but in our experience does not work as well as cross-validation. Theoretical development of a good tuning parameter selector can be of significant importance which we do not investigate here.

Furthermore, one naturally wonders whether a similar RKHS approach can be extended to sufficient dimension reduction such as functional sliced inverse regression (SIR), which was also traditionally based on functional PCA which assumes that the projection direction of interest is well-represented by the basis obtained from functional PCA. It is interesting to see whether the more general framework can lead to better performance in functional SIR.

\section*{Appendix: Proofs}

\noindent\textbf{Proof of Proposition \ref{prp:close}.}
Let $\{\omega_j\}_{j=1}^\infty$ be the eigenfunctions of $K$ corresponding to the eigenvalues $\alpha_1\ge \alpha_2\ge\cdots>0$.  Since $\beta(t,\cdot)\in H_K$, we can write $\beta(t,s)=\sum_j a_j(t)\omega_j(s)$ for some function $a_j$, with $\sum_{j=1}^\infty a_j^2(t)/\alpha_j<\infty$ (pointwise summable in $t$). For any $f=\sum_jf_j\omega_j\in H_K$, $BK^{-1/2}f=\sum_j (f_j/\sqrt{\alpha_j})a_j$. Using this representation, $BK^{-1/2}$ can be natually extended to $L_2$ by defining $BK^{-1/2}f=\sum_j (f_j/\sqrt{\alpha_j})a_j\in L_2$ for any $f\in L_2$. Using Cauchy-Schwartz inequality, this operator is obviously bounded on $L_2$ since the assumption that $\|\beta(t,\cdot)\|_{H_K}\in L_2$ implies $(\sum_j a_j^2/\alpha_j)^{1/2}\in L_2$. \hfill $\Box$\\

\noindent\textbf{Proof of Theorem \ref{thm:upper}.} In the proofs we use $C$ to denote a generic positive constant.
Using $\beta(t,\cdot)=K^{1/2}f(t,\cdot)$, from (\ref{eqn:min2}), 
$$\hat{B}(X^*)=\frac{\sum_i(Y_i\otimes K^{1/2}X_i)}{n}(T_n+\lambda I)^{-1}K^{1/2}X^*,$$
where $I$ is the identity operator, $T_n=K^{1/2}\Gamma_nK^{1/2}$ and $\Gamma_n=\sum_i(X_i\otimes X_i)/n$ is the empirical version of $\Gamma$. Using $Y_i=B(X_i)+\epsilon_i$, and noting that $T_n=\sum_i(K^{1/2}X_i\otimes K^{1/2}X_i)/n$, we have
\begin{eqnarray*}
&&\hat{B}(X^*)-B(X^*)\\
&=&\frac{\sum_i(B(X_i)\otimes K^{1/2}X_i)}{n}(T_n+\lambda I)^{-1}K^{1/2}X^*+\frac{\sum_i(\epsilon_i\otimes K^{1/2}X_i)}{n}(T_n+\lambda I)^{-1}K^{1/2}X^*-B(X^*)\\
&=&BK^{-1/2}\left(T_n(T_n+\lambda I)^{-1}-I\right)K^{1/2}X^*+\frac{\sum_i(\epsilon_i\otimes K^{1/2}X_i)}{n}(T_n+\lambda I)^{-1}K^{1/2}X^*\\
&=&-\lambda BK^{-1/2}(T_n+\lambda I)^{-1}K^{1/2}X^*+\frac{\sum_i(\epsilon_i\otimes K^{1/2}X_i)}{n}(T_n+\lambda I)^{-1}K^{1/2}X^*\\
&=:&A_1+A_2.
\end{eqnarray*}
We first deal with $A_1$. Note $A_1=-\lambda BK^{-1/2}(T+\lambda I)^{-1}K^{1/2}X^*-\lambda BK^{-1/2}(T_n+\lambda I)^{-1}(T-T_n)(T+\lambda I)^{-1}K^{1/2}X^*$.

Using the expansion $K^{1/2}X^*=\sum_j\xi_j^*e_j$, 
\begin{eqnarray}
&&\lambda^2 E^*\|BK^{-1/2}(T+\lambda I)^{-1}K^{1/2}X^*\|^2\nonumber\\
&=& \lambda^2 E^*\left[ \sum_k\langle BK^{-1/2}\sum_j\frac{\xi_j^*}{s_j+\lambda}e_j, e_k\rangle^2\right]\nonumber\\
&=&\lambda^2 \sum_{j,k}\frac{s_j}{(s_j+\lambda)^2} \langle BK^{-1/2}e_j,e_k\rangle\nonumber\\
&\le &\frac{\lambda}{4}\sum_{j,k}\langle BK^{-1/2}e_j,e_k\rangle^2\nonumber\\
&=&\frac{\lambda}{4}\|BK^{-1/2}\|_{HS}^2.\label{eqn:a11}
\end{eqnarray}

Also, writing $\calA=BK^{-1/2}(T_n+\lambda I)^{-1}(T-T_n)(T+\lambda I)^{-1}$ for simplicity of notation,
\begin{eqnarray}
&&\lambda^2 E^*\|BK^{-1/2}(T_n+\lambda I)^{-1}(T-T_n)(T+\lambda I)^{-1}K^{1/2}X^*\|^2\nonumber\\
&=&\lambda^2 E^*\langle \calA K^{1/2}X^*, \calA K^{1/2}X^*\rangle\nonumber\\
&=&\lambda^2 E^*\langle \calA^\T\calA K^{1/2}X^*,  K^{1/2}X^*\rangle\nonumber\\
&=&\lambda^2 {\rm Trace}(\calA^\T\calA T)\nonumber\\
&=& \lambda^2\|\calA T^{1/2}\|_{HS}^2\nonumber\\
&\le& \lambda^2\|BK^{-1/2}\|_{HS}^2\|(T_n+\lambda I)^{-1}(T-T_n)(T+\lambda I)^{-1}T^{1/2}\|^2_{HS}\nonumber\\
&\le& \lambda^2\|BK^{-1/2}\|_{HS}^2\|(T_n+\lambda I)^{-1}\|^2_{op}\|(T-T_n)(T+\lambda I)^{-1}T^{1/2}\|^2_{HS}\nonumber\\
&=&O_p(\|(T-T_n)(T+\lambda I)^{-1}T^{1/2}\|^2_{HS}).\label{eqn:a12}
\end{eqnarray}
We have 
\begin{eqnarray}
&&E\|(T-T_n)(T+\lambda I)^{-1}T^{1/2}\|^2_{HS}\nonumber\\
&=&E\sum_{j,k}\langle(T-T_n)(T+\lambda I)^{-1}T^{1/2}e_j,e_k\rangle^2\nonumber\\
&=&E\sum_{j,k}\langle(T-T_n)\frac{s_j^{1/2}}{(s_j+\lambda)}e_j,e_k\rangle^2.\label{eqn:1}
\end{eqnarray}
Direct calculation reveals that 
\begin{eqnarray*}
&&E\langle (T-T_n)e_j,e_k\rangle^2\\
&=& E \langle s_je_j-\frac{1}{n}\sum_i((\sum_l\xi_{il}e_l)\otimes (\sum_m\xi_{im}e_m))e_j,e_k\rangle^2\\
&=& E\langle s_je_j-\sum_i\frac{\sum_l\xi_{il}\xi_{ij}e_l}{n},e_k\rangle^2\\
&=&E(s_jI\{j=k\}-\frac{\sum_i\xi_{ij}\xi_{ik}}{n})^2\\
&\le& E(\frac{\sum_i\xi_{ij}\xi_{ik}}{n})^2,
\end{eqnarray*}
where the last step used the fact that $E[\xi_{ij}\xi_{ik}]=s_jI\{j=k\}$. Using assumption (A2), we have $E\langle (T-T_n)e_j,e_k\rangle^2\le Cs_js_k/n$, which combined with (\ref{eqn:1}) implies
\begin{equation}\label{eqn:2}
E\|(T-T_n)(T+\lambda I)^{-1}T^{1/2}\|^2_{HS}\le \frac{C}{n}\sum_{j,k}\frac{s_j^2s_k}{(s_j+\lambda)^2}=O\left(\frac{1}{n}\sum_{j}\frac{s_j^2}{(s_j+\lambda)^2}\right).
\end{equation}
(\ref{eqn:a11}),(\ref{eqn:a12}) and (\ref{eqn:2}) together yield $E^*\|A_1\|^2=O_p(\lambda+ \frac{1}{n}\sum_{j}\frac{s_j^2}{(s_j+\lambda)^2} )$.

Now, write $A_2=\frac{\sum_i(\epsilon_i\otimes K^{1/2}X_i)}{n}(T+\lambda I)^{-1}K^{1/2}X^*-\frac{\sum_i(\epsilon_i\otimes K^{1/2}X_i)}{n}(T+\lambda I)^{-1}(T-T_n)(T_n+\lambda I)^{-1}K^{1/2}X^*$. We have
\begin{eqnarray*}
&&E^*\|\frac{\sum_i(\epsilon_i\otimes K^{1/2}X_i)}{n}(T+\lambda I)^{-1}K^{1/2}X^*\|^2\\
&=&E^*\|\frac{1}{n}\sum_i\epsilon_i\langle K^{1/2}X_i,\sum_j \frac{\xi^*_j}{s_j+\lambda}e_j\rangle\|^2,
\end{eqnarray*}
and thus 
\begin{eqnarray*}
&&E\|\frac{\sum_i(\epsilon_i\otimes K^{1/2}X_i)}{n}(T+\lambda I)^{-1}K^{1/2}X^*\|^2\\
&=&\frac{\sigma_\epsilon^2}{n}E\left[\langle K^{1/2}X_1,\sum_j \frac{\xi^*_j}{s_j+\lambda}e_j\rangle^2\right]\\
&=&\frac{\sigma_\epsilon^2}{n}E\sum_j\frac{\xi_{1j}^2\xi_j^{*2}}{(s_j+\lambda)^2}\\
&=&\frac{\sigma_\epsilon^2}{n}\sum_j\frac{s_j^2}{(s_j+\lambda)^2}.
\end{eqnarray*}
where $\sigma_\epsilon^2=E\|\epsilon\|^2$.
Furthermore, denoting $\calC=(T+\lambda I)^{-1}(T-T_n)(T_n+\lambda I)^{-1}$,
\begin{eqnarray*}
&&E\left[\|\frac{\sum_i(\epsilon_i\otimes K^{1/2}X_i)}{n}(T_n+\lambda I)^{-1}(T-T_n)(T+\lambda I)^{-1}K^{1/2}X^*\|^2|X_1,\ldots,X_n\right]\\
&=&\frac{\sigma_\epsilon^2}{n^2}E\left[\sum_i\langle K^{1/2}X_i,(T_n+\lambda I)^{-1}(T-T_n)(T+\lambda I)^{-1}K^{1/2}X^*\rangle^2|X_1,\ldots,X_n\right]\\
&=&\frac{\sigma_\epsilon^2}{n^2}E\left[\sum_i\langle \calC K^{1/2}X_i,K^{1/2}X^*\rangle^2|X_1,\ldots,X_n\right]\\
&=&\frac{\sigma_\epsilon^2}{n^2}\left[\sum_i\langle \calC^\T T\calC K^{1/2}X_i,K^{1/2}X_i\rangle\right]\\
&=&\frac{\sigma_\epsilon^2}{n}{\rm Trace}(\calC^\T T\calC T_n)\\
&=&\frac{\sigma_\epsilon^2}{n}{\rm Trace}(T_n^{1/2}\calC^\T T^{1/2}T^{1/2}\calC T_n^{1/2})\\
&=&\frac{\sigma_\epsilon^2}{n}\|T_n^{1/2}\calC^\T T^{1/2}\|^2_{HS}\\
&=&\frac{\sigma_\epsilon^2}{n}\|T_n^{1/2}(T_n+\lambda I)^{-1}(T-T_n)(T+\lambda I)^{-1}T^{1/2}\|^2_{HS}\\
&\le&\frac{\sigma_\epsilon^2}{n}\|T_n^{1/2}(T_n+\lambda I)^{-1}\|^2_{op}\|(T-T_n)(T+\lambda I)^{-1}T^{1/2}\|^2_{HS}\\
&=&O_p(\frac{1}{n\lambda})\cdot O_p(\frac{1}{n}\sum_{j}\frac{s_j^2}{(s_j+\lambda)^2})\\
&=&o_p(\frac{1}{n}\sum_{j}\frac{s_j^2}{(s_j+\lambda)^2}),
\end{eqnarray*}
where we used (\ref{eqn:2}) and that $n\lambda\rightarrow\infty$.
Thus we have $E^*\|A_2\|^2=O_p(\frac{1}{n}\sum_j\frac{s_j^2}{(s_j+\lambda)^2})$. The theorem is proved by combining the bounds for $E^*\|A_1\|^2$ and $E^*\|A_2\|^2$. \hfill $\Box$\\

\noindent\textbf{Proof of Theorem \ref{thm:lower}.} Our model is $Y(t)=\int \beta(t,s)X(s)ds+\epsilon(t)$. Consider the special case $\beta(t,s)=e_1(t)\otimes \beta(s)$ and $\epsilon(t)=e_1(t)\chi$, where $\beta(s)\in H_K$, $\|\beta\|_{H_K}\le 1$, and $\chi\sim N(0,\sigma^2)$. Then by taking inner products with $e_j$ on both sides of $Y(t)=\int \beta(t,s)X(s)ds+\epsilon(t)$, the model becomes $Y^{(1)}=\int \beta(s)X(s)ds+\chi$, $Y^{(2)}=Y^{(3)}=\cdots=0$ where $Y^{(j)}=\langle Y, e_j\rangle$. Since $\|\int \beta(\cdot,s)X(s)ds\|=|\int \beta(s)X(s)ds|$, the lower bound for the scalar response model provides a lower bound for the functional response model. Thus we can just consider the model with scalar response:
\begin{equation*}
Y_i=\int \beta(s)X_i(s)ds+\chi_i,
\end{equation*}
with $\|\beta\|_{H_K}\le 1$. We need a modification of the proof of Theorem 1 in \cite{caiyuan12} due to the more general assumption on the eigenvalues of $T$. Let $\eta_j=\sqrt{c\lambda_0/(J_0s_j)}$ for some $0<c\le 1$ to be determined later. We apply Theorem 2.5 of \cite{tsybakov09} using the following collection of $2^{J_0}$ functions 
$$f_\theta=\sum_{k=1}^{J_0}\theta_k \eta_k K^{1/2}e_k,$$
where $\theta=(\theta_1,\ldots,\theta_{J_0})\in \{0,1\}^{J_0}$.

First, using that $\|K^{1/2}e_j, K^{1/2}e_k\|_{H_K}=\langle e_j, e_k\rangle=1\{j=k\}$, 
$$\|f_\theta\|_{H_K}^2=\sum_{k=1}^{J_0}\theta_k^2\eta_k^2\le \sum_{k=1}^{J_0}\eta_k^2=\frac{c\lambda_0}{J_0}\sum_{k=1}^{J_0}\frac{1}{s_k}\le \frac{c\lambda_0}{J_0}\frac{J_0}{s_{J_0}}\le c\le 1,$$
 since $s_{J_0}\ge \lambda_0$ by $s_{J_0}=\phi(J_0)$ and the definition $J_0=\lfloor \phi^{-1}(\lambda_0)\rfloor$.

By the Varshamov-Gilbert bound (Lemma 2.9 in \cite{tsybakov09}), there is a subset $\Theta=\{\theta^{0},\ldots,\theta^{N}\}\subset \{0,1\}^{J_0}$ such that $\theta^{0}=(0,\ldots,0)$, $N\ge 2^{J_0/8}$ and $\sum_{k=1}^{J_0}(\theta_k-\theta'_k)^2\ge J_0/8$ whenever $\theta\neq \theta'\in\Theta$.

We have 
\[\|\Gamma^{1/2}(f_{\theta}-f_{\theta'})\|^2=\sum_{k=1}^{J_0}\eta_k^2(\theta_k-\theta'_k)^2s_k\ge\frac{c\lambda_0}{J_0}\frac{J_0}{8}=c\lambda_0/8,\]
verifying condition $(i)$ in Theorem 2.5 of \cite{tsybakov09}. Furthermore, the Kullback-Leibler distance between $P_\theta$ and $P_{\theta'}$ ($P_\theta$ is the joint distribution of training data when $\beta=f_\theta$) can be found to be 
\[K(P_\theta|P_{\theta'})=\frac{n}{2\sigma^2}\sum_{k=1}^{J_0}\eta_k^2(\theta_k-\theta'_k)^2s_k\le \frac{nc\lambda_0}{2\sigma^2},\]
and thus 

\[\frac{1}{N}\sum_{j=1}^NK(P_\theta|P_{\theta'})\le \frac{nc\lambda_0}{2\sigma^2}=\frac{c\phi^{-1}(\lambda_0)}{2\sigma^2}\le \frac{c}{2\sigma^2}(J_0+1)\le \alpha \log N,\]
for some $0<\alpha<1/8$ if $c$ is chosen small enough, verifying condition (ii) in Theorem 2.5 of \cite{tsybakov09}. The lower bound is proved by applying Theorem 2.5 of \cite{tsybakov09}. \hfill $\Box$

\bibliographystyle{jasa}
\bibliography{papers.txt,books.txt}

\end{document}